\documentclass[twocolumn]{aastex61}
\usepackage{longtable}
\usepackage{float}
\usepackage{tablefootnote}

\usepackage{lineno}
\usepackage{amsmath}

\DeclareUnicodeCharacter{2212}{-} 

\submitjournal{ApJ}

\shorttitle{Small-Scale Magnetic Fields of Hyades M Dwarfs}
\shortauthors{Wanderley et al.}

\begin{document}

\title{Small-Scale Magnetic Fields of Hyades M Dwarfs: Evolution with Age from the Pleiades to Field Stars}

\correspondingauthor{Fábio Wanderley}
\email{fabio.wanderley@physics.uu.se}

\author[0000-0003-0697-2209]{Fábio Wanderley}
\affiliation{Department of Physics and Astronomy, Uppsala University, Box 524, S-75120 Uppsala, Sweden}

\author[0000-0001-6476-0576]{Katia Cunha}
\affiliation{Steward Observatory, University of Arizona, 933 North Cherry Avenue, Tucson, AZ 85721-0065, USA}
\affiliation{Observatório Nacional/MCTI, R. Gen. José Cristino, 77,  20921-400, Rio de Janeiro, Brazil}

\author[0000-0003-3061-4591]{Oleg Kochukhov}
\affiliation{Department of Physics and Astronomy, Uppsala University, Box 524, S-75120 Uppsala, Sweden}

\author[0000-0002-0134-2024]{Verne V. Smith}
\affiliation{NSF’s NOIRLab, 950 N. Cherry Ave. Tucson, AZ 85719 USA}

\author[0000-0002-7883-5425]{Diogo Souto}
\affiliation{Departamento de F\'isica, Universidade Federal de Sergipe, Av. Marcelo Deda Chagas, S/N Cep 49.107-230, S\~ao Crist\'ov\~ao, SE, Brazil}

\author[0000-0002-8849-9816]{Lyra Cao}
\affiliation{Department of Physics and Astronomy, Vanderbilt University, Nashville, TN 37235, USA}

\author[0000-0001-6914-7797]{Kevin Covey}
\affiliation{Department of Physics \& Astronomy, Western Washington University, Bellingham, WA, 98225, USA}

\author[0000-0001-9205-2307]{Simone Daflon}
\affiliation{Observatório Nacional/MCTI, R. Gen. José Cristino, 77, 20921-400, Rio de Janeiro, Brazil}

\author[0000-0003-2025-3147]{Steven R. Majewski}
\affiliation{Department of Astronomy, University of Virginia, Charlottesville, VA 22904-4325, USA}

\author[0000-0002-7549-7766]{Marc Pinsonneault}
\affiliation{Department of Astronomy, The Ohio State University, Columbus, OH 43210, USA}

\author[0000-0002-5274-4955]{H. J. Rocha-Pinto}
\affiliation{Universidade Federal do Rio de Janeiro, Observatório do Valongo, Ladeira do Pedro Antônio, 43, Rio de Janeiro, RJ 20080-090, Brazil}

\author[0000-0002-0084-572X]{C. Allende Prieto}
\affiliation{Instituto de Astrofísica de Canarias, Universidad de La Laguna, E-38206 La Laguna, Tenerife, Spain} %affiliation{Departamento de Astrofísica, Universidad de La Laguna, E-38206 La Laguna, Tenerife, Spain}

\begin{abstract}
% 250 words for ApJ

Small-scale magnetic fields are derived for 18 M-dwarf stars in the Hyades open cluster through an analysis of Zeeman broadening in magnetically sensitive lines from the APOGEE spectra, resulting in mean magnetic fields ranging from 0.7 to 4.3 kG. Comparison with results for field stars hosting exoplanets and stars in the Pleiades open cluster from our previous works reveals distinct distributions of mean magnetic fields, reflecting differences in stellar age. 
Pleiades stars exhibit stronger fields ($2965\pm632$ G) and are predominantly in the saturated regime, while the roughly 500 million years older Hyades sample has weaker fields ($1925\pm1028$ G) with a larger dispersion, spanning both saturated and unsaturated regimes. Field stars show substantially weaker fields ($573\pm272$ G) and consistently occupy the unsaturated regime. Using the three samples, we derive the relationship between small-scale magnetic field strength and Rossby number, identifying saturation limits for the Rossby number ($\mathrm{Ro_{SAT}}=0.145$) and mean magnetic field ($\langle\mathrm{B}\rangle_{\rm SAT}=2452$ G).
Within the Hyades sample, we find that, at a given age, cooler stars tend to have stronger magnetic fields. This trend is less apparent in the other samples, due to the broader age distribution of the field stars and magnetic saturation among the Pleiades stars. Comparisons with isochrones indicate positive correlations between mean magnetic field strength and both stellar radius inflation and photospheric spot coverage. We also find positive correlations between magnetic field strength and chromospheric and coronal activity, with stronger fields associated with larger H$\alpha$ equivalent width excess and higher X-ray-to-bolometric luminosity ratios.

\end{abstract}

\section{Introduction}

Understanding the physical properties and evolutionary histories of M dwarf stars requires a detailed understanding of their magnetic fields. Strong magnetic fields play a central role in shaping a range of stellar phenomena, including radius inflation \citep{mullan2001,feiden2015}, starspot formation \citep{chabrier2007}, and enhanced high-energy emission \citep{hawley2014,astudillo2017,newton2017}. The latter is particularly important in the context of exoplanetary systems, as it directly affects the atmospheres and potential habitability of orbiting planets \citep{shkolnik2014}. Strong magnetic fields can also produce RV jitter \citep{ruh2024}, which needs to be accounted for when detecting exoplanets around active stars via the radial velocity methodology. 

The magnetic fields of M dwarf stars show a strong dependence on stellar mass and age. As time passes, the interactions between magnetic fields and stellar winds dissipate the star's angular momentum, reducing its rotational velocity and magnetic field strength. The timescale at which this process happens (i.e., the spin-down timescale) is dependent on the stellar mass. It is expected that for the same age, cooler stars will show, on average, stronger magnetic fields. M-dwarfs are cool stars with highly convective interiors, maintaining strong magnetic fields for longer periods, compared to solar-type stars \citep{newton2016}. This makes them excellent targets for studying stellar magnetic fields.

There are two main techniques that are used to derive stellar magnetic fields. ZDI (Zeeman Doppler Imaging) uses polarization spectra to obtain the star's global magnetic vectorial structure, providing the large-scale magnetic field of the star. However, this method is insensitive to small-scale magnetic structures that dominate the field of the star. The other method to study magnetic fields is by investigating the Zeeman broadening of magnetically sensitive spectral lines (i.e., lines with high effective Landé g-factors). The magnetic field derived using this method is the global mean magnetic field ($\langle \rm B \rangle$), and is generally many times stronger than the one provided by the ZDI methodology. However, it does not provide any information regarding the vectorial structure of the field. In this work, we study global mean magnetic fields using Zeeman broadening analysis.

Many works have derived mean magnetic fields of M-dwarf stars, with values reaching as high as 7-8 kG \citep{shulyak2017_olegreview2,shulyak2019_olegreview3,reiners2022,cristofari2023a,cristofari2023b,cristofari2025,shulyak2011_olegreview1,kochukhov2017_olegreview4, kochukhov2001_olegreview5, kochukhov2009_olegreview6, kochukhov2019_olegreview7, johns-krull2000_olegreview8, afram2009_olegreview9, shulyak2014_olegreview10, saar1985_olegreview11, saar1994_olegreview12, phan-bao_olegreview13, saar1996_olegreview14, johns-krull1996_olegreview15, hahlin2024}. %(see references in Section 4.1). 
These studies focused on M dwarfs in the Galactic field, encompassing stars spanning a broad range of ages. In contrast, stars belonging to the same open cluster are expected to share a common age, allowing their magnetic properties to be compared with those of stars in clusters of different ages. Such comparisons provide a means of tracing the evolution of magnetic fields in low-mass stars over time. 

\citet{wanderley2024} presented the first study to date to derive small-scale mean magnetic fields for a large sample of low-mass stars belonging to the same open cluster. They studied 62 M dwarfs in the Pleiades open cluster, which has an age of $\sim$110--160 Myr (\citealt{gossage2018,stauffer1998,dahm2015,cummings2018}), and found mean magnetic field strengths ranging from $\sim$1.0 to $\sim$4.2 kG. Most of their targets are expected to have experienced relatively little angular-momentum loss and to retain strong magnetic fields, being in the saturated regime. This study found positive correlations between the derived magnetic field strengths and radius inflation, fractional spot coverage, and chromospheric and coronal activity. \citet{wanderley2024b} focused on field M dwarfs known to host planets, finding substantially weaker mean magnetic fields, ranging from $\sim$0.2 to $\sim$1.5 kG, consistent with the older ages of these stars. This sample enabled an investigation of the relationship between mean magnetic field strength and stellar age; however, because the field stars span a broad range of ages, the effects of age cannot be isolated as cleanly as in a comparison between coeval stellar populations.

In this work, we extend this analysis to M dwarfs in the Hyades open cluster, whose age of $\sim$600--700 Myr (\citealt{gossage2018,perryman1998,barrado2004}) is approximately 500 Myr older than that of the Pleiades and substantially younger than the field-star sample. The Hyades provide an intermediate-age population with a well-constrained age, enabling a more systematic investigation of the evolution of magnetic fields with stellar age. In particular, this work provides, for the first time, a direct comparison of Zeeman broadening mean magnetic field measurements for M dwarfs in two open clusters with different ages (for an investigation of the relationship between large-scale magnetic fields derived from ZDI and stellar age, see \citealt{folsom2016,folsom2018,vidotto2014}).

This paper is organized as follows. In Section 2, we present the sample of Hyades M dwarf stars. The methodology to derive atmospheric parameters and mean magnetic fields for our sample is described in Section 3. In Section 4, we discuss our results, which include the relation between mean magnetic fields and stellar ages, radius inflation, starspots, and activity. Finally, Section 5 summarizes the conclusions. 

\section{The Sample and Data}

In this work, we analyze M dwarf stars that are members of the Hyades open cluster and have SDSS APOGEE spectra. The APOGEE survey \citep{majewski2017_apogee} is a high-resolution (R$\sim$22,500) near-infrared spectroscopic survey that operated in the H-band ($\lambda$1.51 $\mu$m - $\lambda$1.69 $\mu$m). The studied spectra were obtained using the cryogenic, multi-fiber (300 fibers) APOGEE-N spectrograph on the 2.5-m telescope at the Apache Point Observatory in New Mexico, USA, %The final data release of the APOGEE survey was published in 2021 with its seventeenth data release (DR17), having observed more than 650,000 stars. The main goal of the survey was to observe red giant stars, however, cooler stars such as M dwarfs were also observed, including Hyades stars observed by 
during the APOGEE BTX (bright time extension) \citep{zasowski2017,beaton2021}. 

%As will be discussed further, our magnetic field determination methodology is only suitable for partially convective M dwarfs. 
The previous work of \citet{wanderley2023} studied radius inflation in a sample of M dwarf members of the Hyades open cluster from an analysis of APOGEE spectra. We selected 15 stars with $T_{\rm eff}>3400$ K from their sample. %They found that, on average, the stars show larger radii than predicted by isochrone models. We derive effective temperatures for their sample, using the same methodology from \citet{wanderley2025}, and selected 15 stars with T$_{\rm eff}>$3400 K. The resulting effective temperatures for this sample are quite similar to the ones from \citet{wanderley2023}, with a mean$\pm$SD $\Delta$ T$_{\rm eff}$ (This Work $-$ \citet{wanderley2023}) of $-$21$\pm$45 K. 
% There are other things that I did in this selection, for example 3 stars have RUWE>1.4 in W2023, but since I just put select 16, I don't think I need to mention it
%We note that while 
The sample of \citet{wanderley2023} was limited to stars with $T_{\rm eff}<3800$ K, here we extended the effective temperature range to $\sim$4000 K and included three M dwarf stars in the threshold between late-K and early-M, using the membership study from \citet{goldman2013}. All stars in our sample are considered to be non-binary according to \citet{douglas2019} and have Gaia DR3 RUWE (Renormalized Unit Weight Error) $<1.4$ \citep{belokurov2020}. Our complete stellar sample is composed of 18 partially convective M to late-K dwarf stars with effective temperatures ranging from roughly 3400 K to 4000 K. We checked the membership status of our Hyades target stars in \citet{CANTATGAUDIN_2020}, and found that all targets were assigned a membership probability of 100$\%$.

\begin{figure*}
\begin{center}  \includegraphics[angle=0,width=1\linewidth,clip]{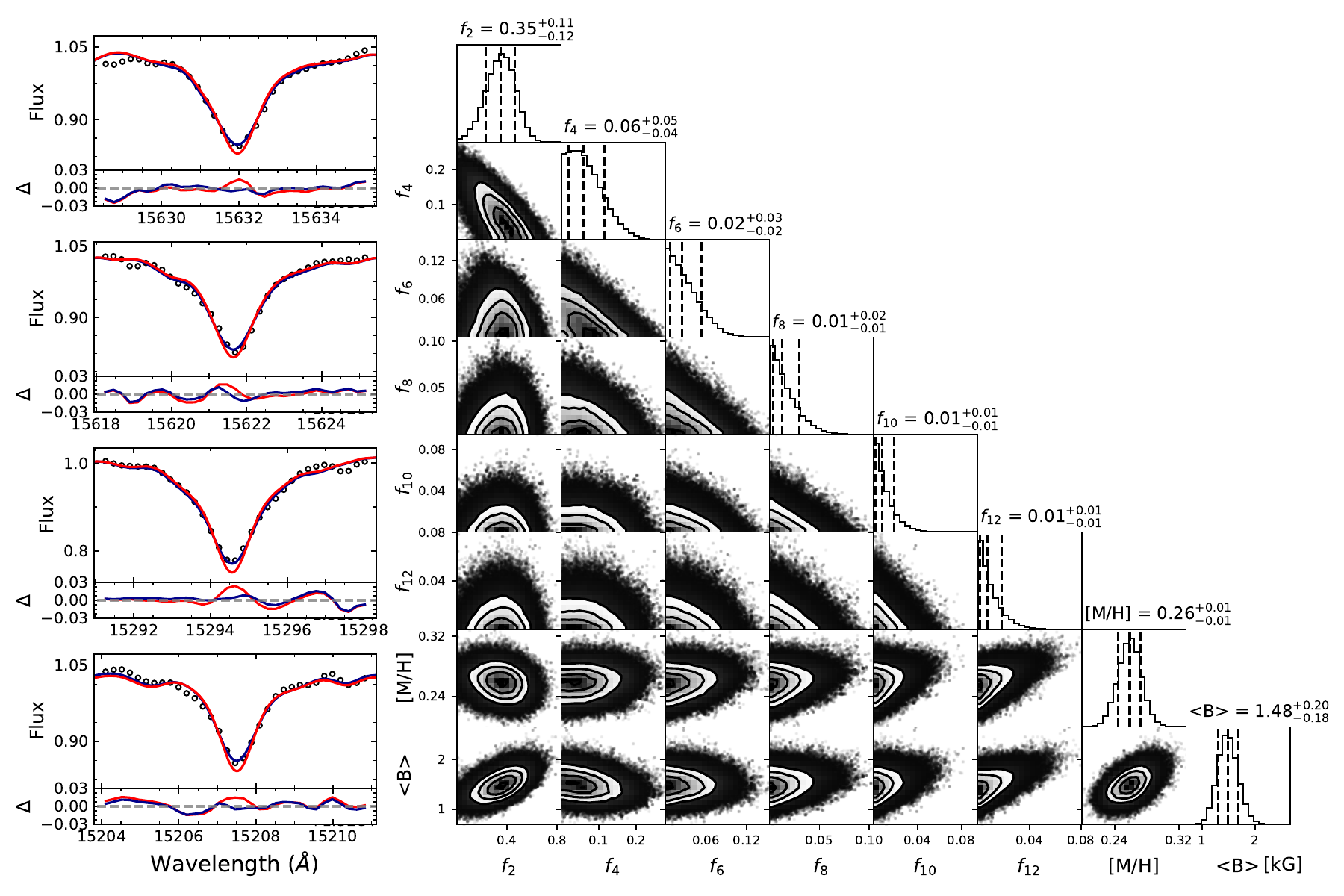}
\caption{%Here we present our
Mean magnetic field results for the M dwarf star 2M04295572+1654506. The upper-left panels show the four Fe I lines used in the modeling. Open symbols are the observed APOGEE spectrum, red lines are synthetic profiles computed without considering magnetic fields, and blue lines are our best fits associated with our derived magnetic field strengths. The lower-left panels show the difference between the observed spectrum and each synthesis. In the right panel, we present the corner plot from the MCMC analysis for this star, showing the posterior distributions of the magnetic filling factors, metallicity, and mean magnetic field strength. We also indicate the median values and corresponding uncertainties, derived from the 16th and 84th percentiles of the posterior distributions, for each parameter.}
\end{center}
\label{magexample}
\end{figure*}

\section{Methodology}

\subsection{Stellar Parameters}

In this study, we first derived the effective temperatures ($T_{\rm eff}$), surface gravities ($\log{g}$), and projected rotational velocities ($v\sin{i}$) of the Hyades M-dwarf stars, as these parameters are inputs for calculating mean magnetic fields. We adopted the methodology of \citet{wanderley2025}, which employs an MCMC approach together with the radiative transfer code Turbospectrum \citep{plez2012_turbospectrum}, MARCS atmosphere models \citep{gustafsson2008_marcs}, and the APOGEE DR17 line list \citep{smith2021}. Effective temperatures and surface gravities, along with oxygen abundances, were obtained using spectral windows containing molecular diagnostic lines of water and OH given in \citet{wanderley2025b}, exploring their distinct sensitivities to $T_{\rm eff}$ and $\log{g}$, as discussed previously in \citet{souto2018,souto2020}. Metallicities were also derived at the same time from global fits of the observed spectra to synthetic spectra with varying metallicities. Stellar projected rotational velocities were determined by modeling the OH lines, which are not very sensitive to magnetic fields.

Stellar radii were derived using the Stefan-Boltzmann equation, applying our derived effective temperatures, along with stellar luminosities. To derive stellar luminosities, we employed the relation from \citet{mamajek2015} between stellar luminosity and bolometric magnitude. These magnitudes were calculated using stellar distances from \citet{bailerjones2021} and 2MASS J-band bolometric corrections, which were derived using photometric calibrations from \citet{mann2015,mann2016}. The uncertainties of the derived stellar parameters are $\pm$100 K for $T_{\rm eff}$ and $\pm$0.2 dex for $\log{g}$ (see discussion in \citealt{wanderley2025}). Regarding $v\sin{i}$, an average uncertainty of $\pm 1~\text{km s}^{-1}$ was adopted, derived from the posterior distributions obtained when modeling the OH lines. Comparing the 15 overlapping stars between this study and \citet{wanderley2023}, our derived parameters show strong consistency with the values reported by this previous study. We find mean differences (This work $-$ \citealt{wanderley2023}) in $T_{\rm eff}$, $\log{g}$, and $R_{*}$ of $-21 \pm 45$,K, $+0.02 \pm 0.04$,dex, and $+0.005 \pm 0.012$ $R_{\odot}$, respectively.

\subsection{Mean Magnetic Fields}
%In this study, we derive mean magnetic fields for 18 M dwarfs observed by the APOGEE survey that are members of the Hyades open cluster. 
The determination of mean magnetic fields for our sample used the same methodology and grid of synthetic spectra presented in \citet{wanderley2024}. The mean magnetic field $\langle \rm B \rangle$ of a star can be obtained by modeling its magnetic filling factors, which characterize the resulting distribution of magnetic field strengths across the stellar surface. This is particularly important for partially convective M dwarfs, whose magnetic field topologies tend to be more complex than those of fully convective stars \citep{donati2008,reiners2009}.
% Due to being fully-convective, fully-convective stars have a more homogeneous magnetic field distribution, and are therefore less spotted.
A hypothetical star whose entire surface is covered by a 2 kG magnetic field has the same $\langle \rm B \rangle$ as a star with 50$\%$ of its surface covered by a 4 kG magnetic field and the remaining 50$\%$ devoid of magnetic fields. However, the effect of the magnetic field on sensitive lines is remarkably different when comparing the two cases.

%for the Pleiades open cluster. This library of synthetic spectra was computed with the radiative transfer code Synmast \citep{kochukhov2010_synmast}. %along with MARCS atmospheric models \citep{gustafsson2008_marcs}, and the APOGEE DR17 line list \citep{smith2021}.
As previously, the same four Fe I lines that are sensitive to magnetic fields, having large effective Landé g-factors, were employed in the analysis: $\lambda$15207.526 \AA, $\lambda$15294.56 \AA, $\lambda$15621.654 \AA, and $\lambda$15631.948 \AA. 
%to produce a library of grant
%having synthetic spectra varying T$_{\rm eff}$ (in 100 K steps), $\log{g}$ (in 0.5 dex steps), [M/H] (in 0.25 dex steps), and $<$B$>$ (in 2000 G steps). 
For each star, %from our sample, 
we interpolated within the synthetic grid adopting the derived $T_{\rm eff}$ and $\log{g}$. 
We also convolved each synthetic spectrum with the instrumental line spread function (LSF), modeled as a Gaussian profile, and with the derived $v\sin{i}$, using a rotational profile.
%The effective temperatures, surface gravities, and projected rotational velocities adopted in the calculations of the mean magnetic fields were derived using the same methodology as \citet{wanderley2025}. For details, we refer to this previous study.
%The mean magnetic field given in Gauss units is defined by the equation below, where f$_{n}$ are the magnetic filling factors. 
%The magnetic filling factor associated with a field of 0 G is given by 1$-\sum f_{n}$:
%\begin{equation}
%    <B>= \sum_{n} f_{n} \times 1000n \, \, , \, \,
%    n=[2,4,6,8,10,12]
%    \label{eqb}    
%\end{equation}
We employed the Python code emcee \citep{emcee2013} to produce a posterior distribution that simultaneously varies metallicity and 6 filling factors (2--12 kG in a 2 kG step), to fit the four selected Fe I lines. %In the MCMC procedure, we combine synthesis with different magnetic fields, multiplying each synthesis by its respective filling factor. 
%This combined synthesis is then compared with the observed spectra using our MCMC methodology, which simultaneously varies the magnetic filling factors and the metallicity. One possible way to derive the mean magnetic field from these results would be to adopt the median value of the posterior distribution of each filling factor. However, due to the degeneracy among the magnetic filling factors — meaning that different combinations of parameters can produce similarly good fits — we instead follow the same procedure adopted by \citet{wanderley2024} to derive the mean magnetic fields.
For each entry of the posterior distribution, we %apply Equation \ref{eqb}, 
derive a corresponding magnetic field strength. The adopted $\langle \rm B \rangle$ is given by the median of this distribution, while the lower and upper uncertainties are given by the offset between the median and, respectively, the 16th and 84th percentiles. 

%\begin{equation}
%    L_{*}=(3.0128 \times 10^{35} ) 10^{-0.4M_{\rm bol}}
%    \label{eqmamajek}    
%\end{equation}

Figure \ref{magexample} shows the MCMC results for the star 2M04295572+1654506. The upper-left panels show the fits obtained to the four magnetic Fe I lines. The observed spectrum is shown as open circles, while the red and blue lines represent the spectral syntheses obtained without magnetic fields and those corresponding to the derived magnetic field strength of 1.48$^{+0.20}_{-0.18}$ kG, respectively. The lower-left panels show the difference between the observed spectrum and each synthesis. The right panel is the corner plot, which shows the relation between the parameters (i.e., magnetic filling factors, metallicity, and mean magnetic field), along with the median and uncertainties of their posterior distributions.

\begin{figure}
\begin{center}  \includegraphics[angle=0,width=1\linewidth,clip]{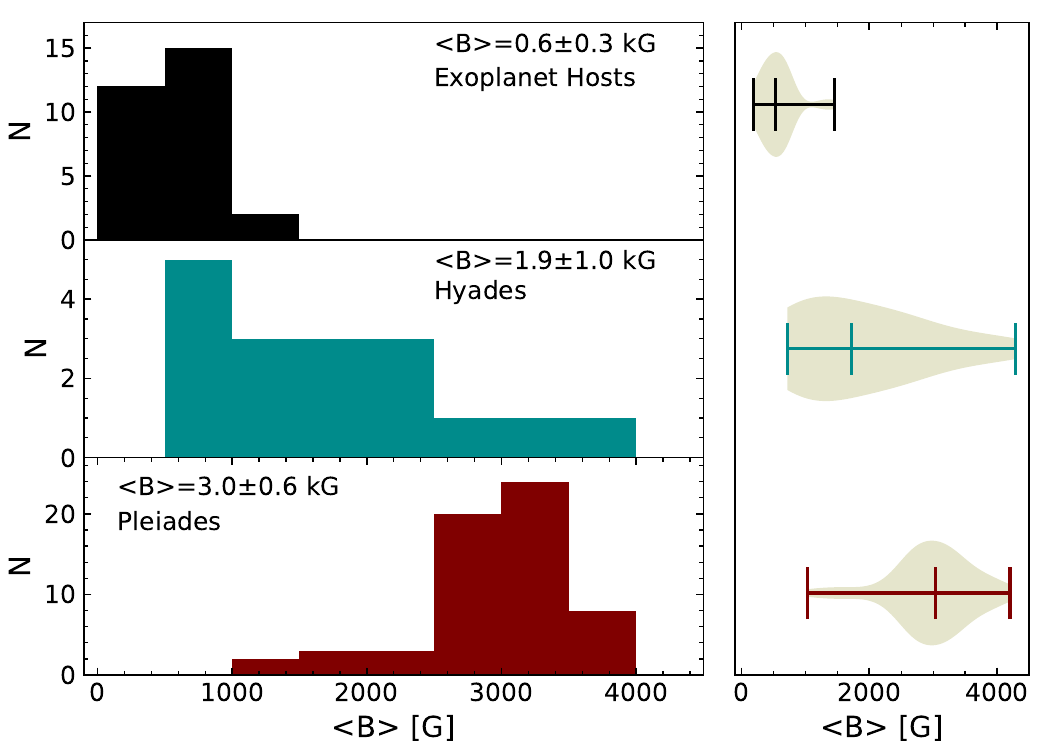}
\caption{Left panels: Mean magnetic fields histograms for M dwarf stars. The top panel shows results for field stars that are known to host planets from \citet{wanderley2024b}, the middle panel shows results for Hyades stars from this work, and the bottom panel shows results for Pleiades stars from \citet{wanderley2024}. We also include the mean$\pm$SD (standard deviation) for each sample. Right panel: ``violin'' plots showing the range, median, and dispersion of each dataset.} \end{center}
\label{hists}
\end{figure}

%\begin{figure}
%\begin{center}  \includegraphics[angle=0,width=1\linewidth,clip]{lit.pdf}
%\caption{Mean magnetic fields as a function of rotational periods. Results for Hyades M dwarfs from this work are represented as black circles. Green squares, maroon x’s, and cyan diamonds are data for field stars from respectively \citet{shulyak2017_olegreview2,shulyak2019_olegreview3}, \citet{reiners2022}, and \citet{cristofari2023a,cristofari2023b,cristofari2025}. Orange triangles show results for field stars from other works; see text for references. The gray dashed lines indicate our determinations of P$_{\rm rot, SAT}$ and $\langle \mathrm{B} \rangle_{\rm SAT}$, while the black lines represent our linear regression fits for the saturated and unsaturated regimes.} 
%\end{center}
%\label{lit}
%\end{figure}

\section{Discussion}

\subsection{Mean Magnetic Field Distributions}

The distribution of mean magnetic fields obtained for the studied Hyades M dwarf sample is presented in Figure \ref{hists}. The left panels of this figure show histograms of the mean magnetic field strengths for three different M-dwarf stellar samples derived by our group. 
The top, middle, and bottom panels of the figure show, respectively: field M dwarf stars known to host planets (from \citealt{wanderley2024b}), Hyades M dwarf stars (this work), and Pleiades M dwarf stars (from \citealt{wanderley2024}). The right panel of Figure \ref{hists} shows ``violin'' plots for each sample, depicting the range, median, and dispersion of each data set. All three samples are composed of M dwarfs within similar effective temperature ranges, but one variable that differentiates the three samples is stellar age. For the same stellar mass, older stars are expected to have dissipated more of their original angular momentum, having lower magnetic field strengths \citep{kawaler1988}. 
%We note that, while in this work we use the same methodology employed by \citet{wanderley2024}, \citet{wanderley2024b} used a very similar approach. However, it considered filling factors up to 6 kG instead of up to 12 kG. 

There is clearly a distinct behavior between the magnetic field distributions for the three populations of M dwarfs shown in Figure \ref{hists}. %which is related to their stellar ages. 
The Pleiades open cluster is very young, with an age of $\sim$120 Myr, and its stars have not had much time to spin down and lose their angular momentum. The Hyades open cluster is approximately six times older, with an age of $\sim$600 Myr. Its stars have had more time to lose their primordial magnetic fields. The mean$\pm$SD $\langle \rm B \rangle$ obtained for the Pleiades and Hyades M dwarfs are, respectively, $2965\pm632$ G and $1925\pm1028$ G. The Pleiades M dwarf stars, which are highly active, have mean magnetic fields generally exceeding 2 kG, whereas the Hyades M dwarf sample exhibits a much wider spread in mean magnetic fields, being concentrated at significantly lower $<$B$>$ values. The field M dwarf stars' $<$B$>$ is even lower and shows significantly less scatter than the Hyades, with an average of $573\pm272$ G, a result in line with the general expectation for stars that are much older. To estimate ages for the 28 field exoplanet host M dwarfs with available peculiar velocity data, we applied the Bayesian kinematic age method of \citet{almeida2018}. This method converts each star's measured $U,V,W$ velocities into a posterior age PDF calibrated on the isochronal ages and peculiar velocities from the Geneva–Copenhagen Survey (GCS; \citealt{nordstrom2004}). The method reproduces isochronal ages both individually (within reported uncertainties) and in ensemble. Using velocities from \citet{laverny2025}, we obtain kinematic ages between 1.8 and 11.6 Gyr (Table \ref{fulldata}). Despite large uncertainties in individual age estimates ($\sim 3\text{ Gyr}$), the sample mean age of $4.6 \pm 1.9 \text{ Gyr}$ indicates that our sample of M-dwarf exoplanet hosts represents an older population than the Hyades and Pleiades, consistent with weaker overall magnetic fields.

\subsection{Mean Magnetic Fields, Stellar Rotation and Rossby Numbers}
It is well established in the literature that there exists a strong dependence between stellar magnetic fields and rotational periods \citep{skumanich1972}, with stars of the same mass that rotate faster presenting, on average, higher magnetic fields than slower-rotating stars. The relation between magnetic fields and rotational periods is often separated into two different regimes. Stars that are slow rotators are located in the unsaturated regime, a location in the $\log{\rm P_{\rm rot}}$ versus $\log{\langle \rm B \rangle}$ plane where there is a strong dependence between both variables. Rapid rotators are located in the saturated regime, a portion of the plane where the relation between $\log{\rm P_{\rm rot}}$ and $\log{\langle \rm B \rangle}$ still exists but is less prominent. Another important quantity that relates to magnetic fields is the Rossby number, which is inversely correlated with magnetic fields and is given by the ratio between the rotational period and the convective turnover time (i.e., the timescale for a convective cell of gas to traverse the convective zone of the star). For the same rotational period, lower-mass main-sequence stars are expected to exhibit stronger magnetic fields due to their longer spin-down timescales. Therefore, Rossby numbers provide a better parameter for comparing magnetic fields across stars of different masses and for separating stars into saturated and unsaturated regimes (\citealt{noyes1984}; See also \citet{pizzolato2003}, who investigated the relationship between stellar activity and Rossby number for late-type stars spanning a range of masses).
% pizzolato2003 compared LX_Lbol for stars with different masses, and found that the pattern is the same

 To derive Rossby numbers, we estimated convective turnover times ($\tau$, in days) using the equation $\tau=12.3/\rm L_{*}$ from \citet{reiners2022}, where L$_{*}$ is the stellar luminosity in solar units. We collected rotational periods for our Hyades sample from   \citet{hartman2011,delorme2011,douglas2016,douglas2019}, using the mean of the measurements when more than one was available, noting that four stars from our sample do not have rotational period measurements available. The derived magnetic fields, along with effective temperatures, projected rotational velocities, rotational periods, Rossby numbers, and stellar radii, are presented in Table \ref{fulldata}. 

\subsubsection{The Boundary Between the Saturated and Unsaturated Regimes}
\begin{figure*}
\begin{center}  
\includegraphics[angle=0,width=0.8\linewidth,clip]{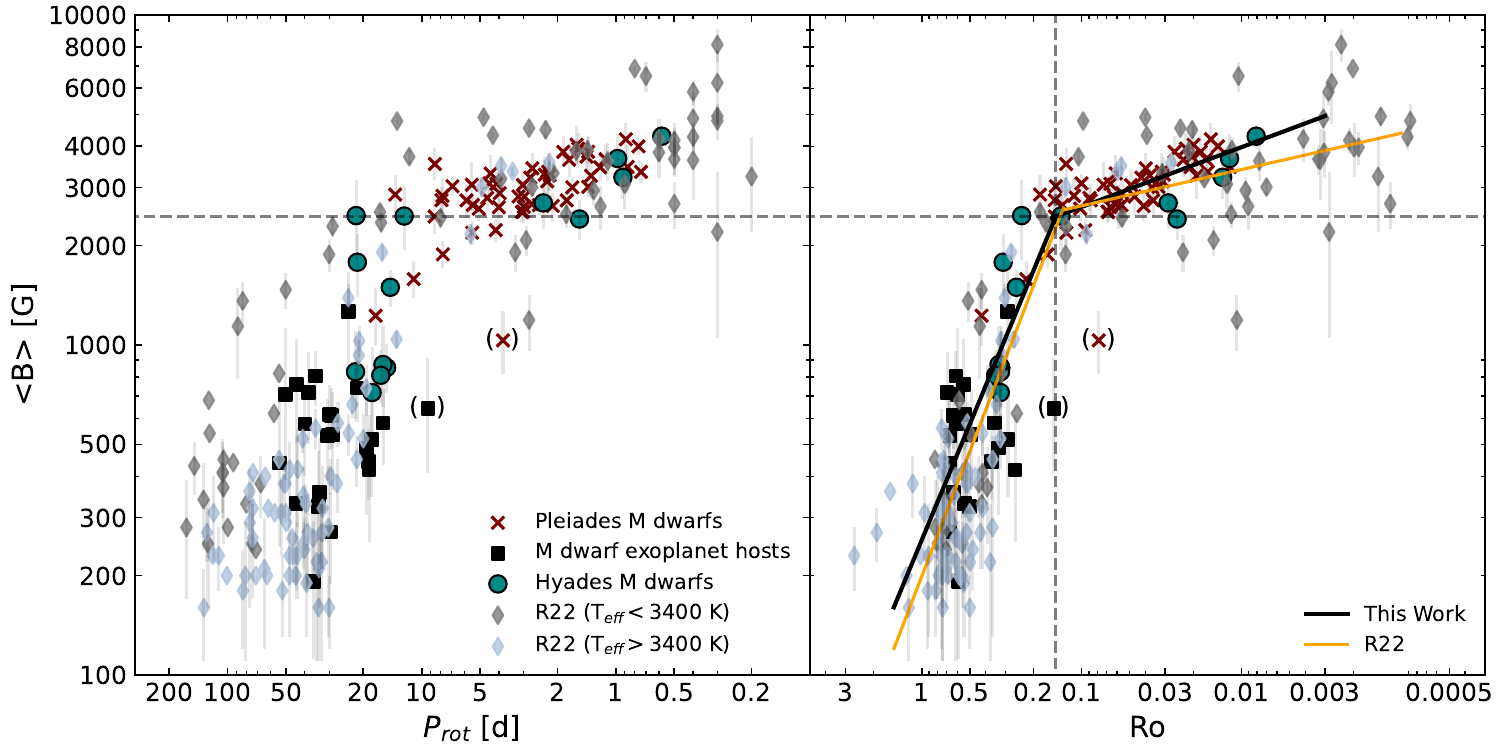}
\caption{Mean magnetic field strengths as a function of rotational period (left panel) and Rossby number (right panel). Results for Hyades M dwarfs from this work are shown as green circles, while results for Pleiades M dwarfs \citep{wanderley2024} and for field stars known to host planets \citep{wanderley2024b} are shown as maroon xs and black squares, respectively. Gray and light blue diamonds show the results of \citet{reiners2022} for stars with $T_{\rm eff}<3400$ K and $T_{\rm eff}>3400$ K, respectively. In the right panel, the black lines represent the $\log{\langle \rm B \rangle}$--$\log{\mathrm{Ro}}$ relations derived from the combined three samples, while the orange lines indicate the relations from \citet{reiners2022}. Dashed gray lines mark the saturation limits: the horizontal lines in both panels indicate $\langle \rm B\rangle_{\rm SAT}$, while the vertical line in the right panel indicates the Ro$_{\rm SAT}$ position.}
\end{center}
\label{mag}
\end{figure*}

Figure \ref{mag} shows mean magnetic fields as a function of rotational periods (in days, left panel) and Rossby numbers (right panel). Our determinations for Hyades M dwarfs are shown as green circles, results from \citet{wanderley2024} for Pleiades M dwarf stars are shown as maroon xs, and field stars from \citet{wanderley2024b} as black squares. The results shown in this figure allow us to estimate the Rossby number and mean magnetic field saturation limits, $\mathrm{Ro_{SAT}}$ and $\mathrm{\langle B \rangle_{SAT}}$, respectively, defining the boundaries between the saturated and unsaturated magnetic activity regimes for our sample. 
The previous study by \citet{wright2011} investigated the rotation--activity relation across FGKM spectral types by examining the relationship between Rossby number and the ratio of X-ray to bolometric luminosity, finding $\mathrm{Ro_{SAT}}=0.13\pm0.02$.
%The gray line in the right panel represents the relation between $\log{\langle \rm B \rangle}$ and $\log{Ro}$ for Pleiades stars from \citet{wanderley2024} with Ro$<0.13$. The orange lines show the relations from \citet{reiners2022} for Ro$<0.13$ (solid line) and Ro$>0.13$ (dashed line), where Ro$=0.13$ is the thresluminosityhold adopted in these works to separate the saturated and unsaturated magnetic-field regimes. 

\begin{figure}
\begin{center}  
\includegraphics[angle=0,width=1\linewidth,clip]{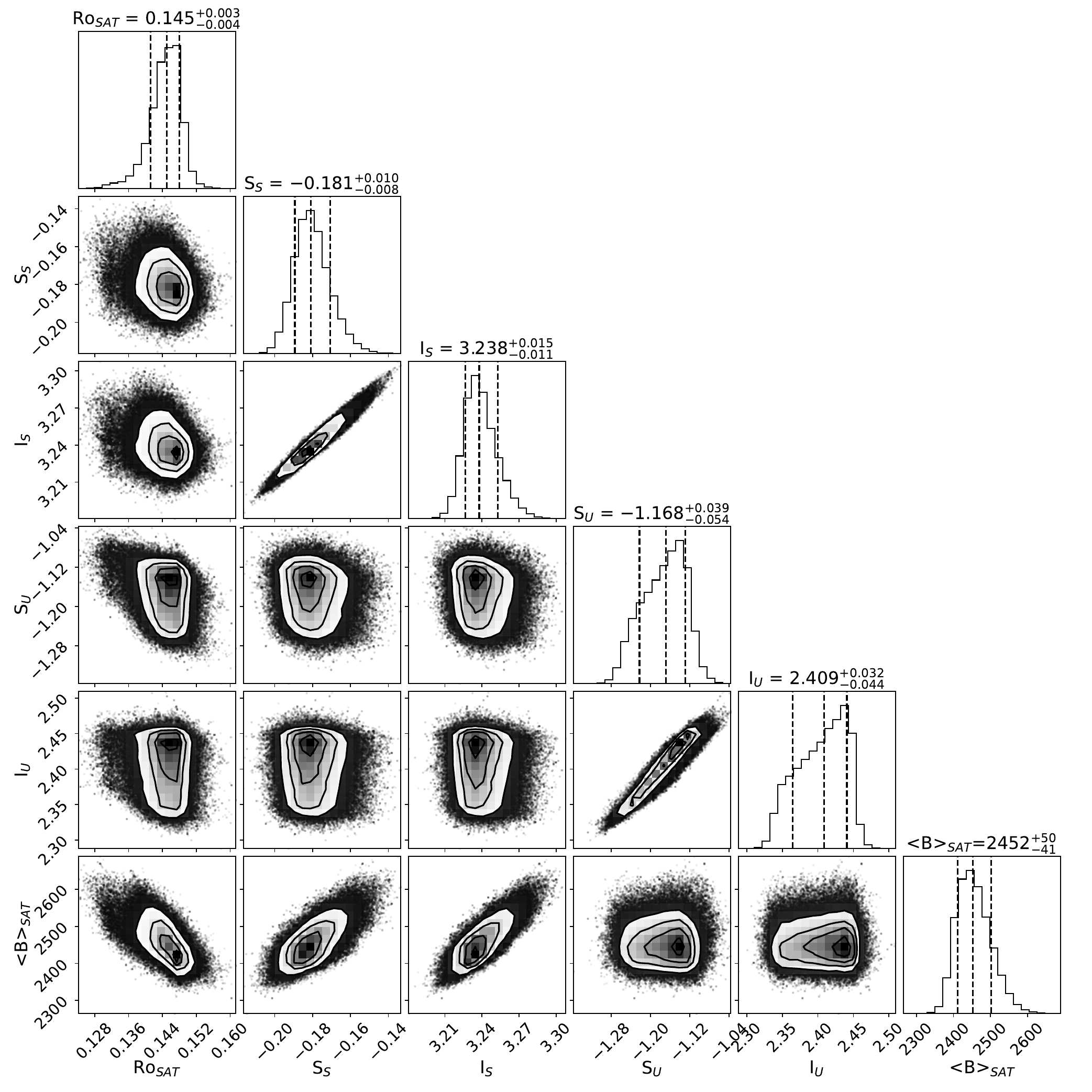}
\caption{Corner plot showing the posterior distributions of the Rossby number and mean magnetic field saturation limits, $\mathrm{Ro_{SAT}}$ and $\langle \mathrm{B}\rangle_{\rm SAT}$, respectively, together with the posterior distributions of the slopes and intercepts of the linear relations between $\log\langle \mathrm{B} \rangle$ and $\log\mathrm{Ro}$ in the saturated and unsaturated regimes. S$_{S}$ and I$_{S}$ denote the slope and intercept of the saturated regime, respectively, while S$_{U}$ and I$_{U}$ denote those of the unsaturated regime. The median values and corresponding uncertainties, derived from the 16th and 84th percentiles of the posterior distributions, are indicated for each parameter.}
\end{center}
\label{cornerrossby}
\end{figure}

To determine $\mathrm{Ro_{SAT}}$ and $\mathrm{\langle B \rangle_{SAT}}$, we used all stars with available rotational period measurements, excluding two outliers that exhibit weaker magnetic fields than expected for their Rossby numbers (marked with parentheses in Figure \ref{mag}). The resulting sample comprises 88 stars, including members of the Hyades and Pleiades clusters, as well as field stars. We assumed that the relationship between $\log{\langle\mathrm{B}\rangle}$ and $\log{\mathrm{Ro}}$ is described by two distinct linear relations: one for the saturated regime ($\mathrm{Ro}\leq\mathrm{Ro_{SAT}}$) and another for the unsaturated regime ($\mathrm{Ro}>\mathrm{Ro_{SAT}}$). We used an MCMC approach to simultaneously determine $\mathrm{Ro_{SAT}}$, $\mathrm{\langle \mathrm{B} \rangle_{SAT}}$, and the slopes and intercepts of the two relations. In the MCMC analysis, we adopted a prior requiring the two linear relations to approximately intersect at the Rossby number saturation limit. For each posterior sample, we calculated the corresponding magnetic field saturation limit. The adopted parameter values are the medians of their respective posterior distributions, with uncertainties given by the difference between the median and the 16th and 84th percentiles. We obtain $\mathrm{Ro}_{\mathrm{SAT}}=0.145$, which is slightly larger than the value reported by \citet{wright2011}, and a mean magnetic field saturation limit of 2452 G. The posterior distributions and parameter correlations are presented in Figure \ref{cornerrossby}, where S$_{S}$ and I$_{S}$ denote the slope and intercept associated with the saturated regime, respectively, while S$_{U}$ and I$_{U}$ denote those associated with the unsaturated regime. The resulting Rossby number and mean magnetic field saturation limits, together with the linear relations for each regime, are presented in the equations below.
\begin{equation}
\label{linearfits0}
\mathrm{\langle B \rangle_{SAT}} = 2452^{+50}_{-41} \, \mathrm{G} 
\end{equation}
\begin{equation}
\label{linearfits1}
\mathrm{Ro_{SAT}} = 0.145^{+0.003}_{-0.004}
\end{equation}
\begin{equation}
\label{linearfits2}
 \log{\langle \rm B \rangle}_{\mathrm{Ro}<\mathrm{Ro_{SAT}}} = 3.238^{+0.015}_{-0.011}-0.181^{+0.010}_{-0.008} \times \log{\mathrm{Ro}}
\end{equation}
\begin{equation}
\label{linearfits3}
 \log{\langle \rm B \rangle}_{\mathrm{Ro}>\mathrm{Ro_{SAT}}} = 2.409^{+0.032}_{-0.044}-1.168^{+0.039}_{-0.054} \times \log{\mathrm{Ro}}
\end{equation}

%\begin{subequations}
%\begin{align}
%    \mathrm{\langle B \rangle_{SAT}} &= 2452^{+50}_{-41} \, \mathrm{G} \label{eq:one_a}\\
%    \mathrm{Ro_{SAT}} &= 0.145^{+0.003}_{-0.004} \label{eq:one_b}\\
%    \log{\langle \rm B \rangle}_{\mathrm{Ro}<\mathrm{Ro_{SAT}}} &= 3.238^{+0.015}_{-0.011}-0.181^{+0.010}_{-0.008} \times \log{\mathrm{Ro}} \label{eq:one_c}\\
%    \log{\langle \rm B \rangle}_{\mathrm{Ro}>\mathrm{Ro_{SAT}}} &= 2.409^{+0.032}_{-0.044}-1.168^{+0.039}_{-0.054} \times \log{\mathrm{Ro}} \label{eq:one_c}
%\end{align}
%\end{subequations}

%To derive the magnetic field saturation limit, $\langle \rm B \rangle_{\rm SAT}$, we investigated the saturated and unsaturated linear regression fits in the $\log{\langle \rm B \rangle}$–$\log{\mathrm{Ro}}$ plane. The corresponding mean magnetic field saturation limit, $\langle \rm B \rangle_{\rm SAT}$, is the magnetic field strength at ${\mathrm{Ro}_{\rm SAT}}$, which corresponds to $\langle \rm B \rangle_{\rm SAT} = 2417$ G. To determine the corresponding saturation limit in rotational-period space, we applied the same procedure to the $\log{P_{\rm rot}}$--$\log{\langle \rm B \rangle}$ relation, obtaining $P_{\rm rot,SAT} = 6.43$ days. 
The derived linear fits are shown in black lines in the right panel of Figure \ref{mag}, along with the relations from \citet{reiners2022} in orange. The gray dashed line indicates the location of ${\mathrm{Ro}_{\rm SAT}}$, and $\langle \rm B \rangle_{\rm SAT}$.
Using only the results for the Pleiades sample, the previous work of \citet{wanderley2024} also derived a linear relation between $\log{\langle \rm B \rangle}$ and $\log{\mathrm{Ro}}$ for the saturated regime. As expected, given the substantial overlap between the two datasets, this relation closely agrees with our relation in Equation \ref{linearfits3}.
%Their relation closely overlaps with our relation in the saturated regime. %relation and is therefore omitted from the figure for clarity. 
It is important to note that, as indicated by our linear fit, the saturated regime is not completely flat, as $\log{\langle \rm B \rangle}$ still exhibits a negative dependence on $\log{\mathrm{Ro}}$. 
Our results show that while the Pleiades M dwarf stars are mostly in the saturated regime, the Hyades M dwarfs are in both saturated and unsaturated regimes, in line with the higher standard deviation in the mean magnetic fields of the Hyades cluster, when compared to the Pleiades cluster. The field M dwarf stars from our sample have much higher rotational periods and Rossby numbers, all of them populating the unsaturated regime. 

For stars with a convective envelope, angular momentum lost through the interaction between stellar winds and magnetic fields can initially be replenished by the exchange of angular momentum from the core to the envelope. The rotational coupling timescale, $\tau_{\rm cpl}$, characterizes the timescale over which the core and envelope become rotationally coupled. \citet{spada2026} derived relations between $\tau_{\rm cpl}$ and stellar mass for stars with masses between 0.4 and 1.25 M${\odot}$. Applying these relations to an adopted Pleiades age of 120 Myr, we find that only stars with masses above 0.79 M${\odot}$ are expected to be rotationally coupled. 
%Applying the relations from \citet{spada2026} to 
For a mass range comparable to that of our complete sample ($\sim$ 0.4 -- 0.6 M$_{\odot}$), we obtain rotational coupling timescales ranging from 565 to 1187 Myr. In this context, our Pleiades M dwarf stars likely still retain a significant reservoir of angular momentum in their cores.

For the Hyades, adopting an age of 650 Myr, the corresponding mass limit decreases to 0.56 M$_{\odot}$. This suggests that some of the stars in our sample may already be rotationally coupled. We note that, although both stars in the saturated and unsaturated regimes are present below this mass limit, all five stars with masses above this limit are in the unsaturated regime, with Rossby numbers ranging from 0.25--0.35, which are greater than $\mathrm{Ro}_{\rm SAT}$, and magnetic field strengths of 0.7--1.5 kG, which are below $\langle \mathrm{B} \rangle_{\rm SAT}$. The youngest star in our field-star sample has an estimated age of 1.8 Gyr. At this age, extrapolating the relations of \citet{spada2026} yields a coupling mass below the fully convective boundary, where the relations are not physically applicable. Nevertheless, the extrapolation suggests that all of the field stars from our sample should be rotationally coupled by this stage, with angular-momentum loss playing a greater role in their continued spin-down.

%\subsubsection{VVV}
In Figure \ref{mag} we also show for comparison the results for field M dwarfs from \citet{reiners2022} and mark as gray diamonds their subsample of fully convective stars (3000 K $\lesssim T_{\rm eff}<3400$ K) and light blue diamonds their subsample of partially convective stars ($3400 <T_{\rm eff}<4000$ K), keeping in mind that our sample (Hyades, Pleiades and exoplanet hosts) is composed of partially convective M dwarf stars. As can be seen in this figure, our sample of M-dwarf exoplanet hosts (black squares) shows approximately the same $\log{\langle \rm B \rangle}$--$\log{P_{\rm rot}}$ and $\log{\langle \rm B \rangle}$--$\log{\mathrm{Ro}}$ relations as the partially convective stars from \citet{reiners2022} in the unsaturated regime.
The fully convective unsaturated stars from \citet{reiners2022} also follow this overall similar behavior as a function of Rossby number. However, there is a clear distinct behavior as a function of rotational periods.
It is notable that for a given rotational period, fully convective unsaturated stars from \citet{reiners2022} present higher $\langle \mathrm{B} \rangle$ values compared to our field star sample. 
This distinct pattern illustrates that, differently from the behavior with rotational period, Rossby numbers account for the effect of stellar mass on magnetic field generation and amplification due to lower-mass stars presenting deeper convective envelopes, higher convective turnover times, and spin-down timescales. This also demonstrates that, while we can define a global Rossby number saturation limit, the corresponding saturation threshold in rotational period varies with stellar mass. This result is analogous to \citet{wright2018}, who showed that, although unsaturated partially and fully convective stars follow the same relationship between activity, as measured by L$_{\rm X}$/L$_{\rm bol}$, and Rossby number, they follow different relationships when activity is considered as a function of rotational period, with unsaturated fully convective stars exhibiting longer rotational periods at a given activity level.

In the saturated regime (see Figure \ref{mag}), our sample consists only of young partially convective M dwarfs in open clusters (ages of $\sim$120 Myr for the Pleiades and $\sim$600 Myr for the Hyades), whereas the sample from \citet{reiners2022} is composed predominantly of fully convective field M dwarfs. Although the individual ages of the stars in \citet{reiners2022} are not known, their field-star nature indicates that they are likely substantially older than the Pleiades and Hyades stars, with ages likely extending into the Gyr regime. Nevertheless, some stars in their sample reach lower Rossby numbers and rotation periods, as well as higher mean magnetic field strengths than those in our open-cluster samples. This is likely related to their deeper convective envelopes, which may support stronger magnetic fields even at older ages.

\subsubsection{Mean magnetic fields and Effective Temperatures}
\begin{figure}
\begin{center}  \includegraphics[angle=0,width=1\linewidth,clip]{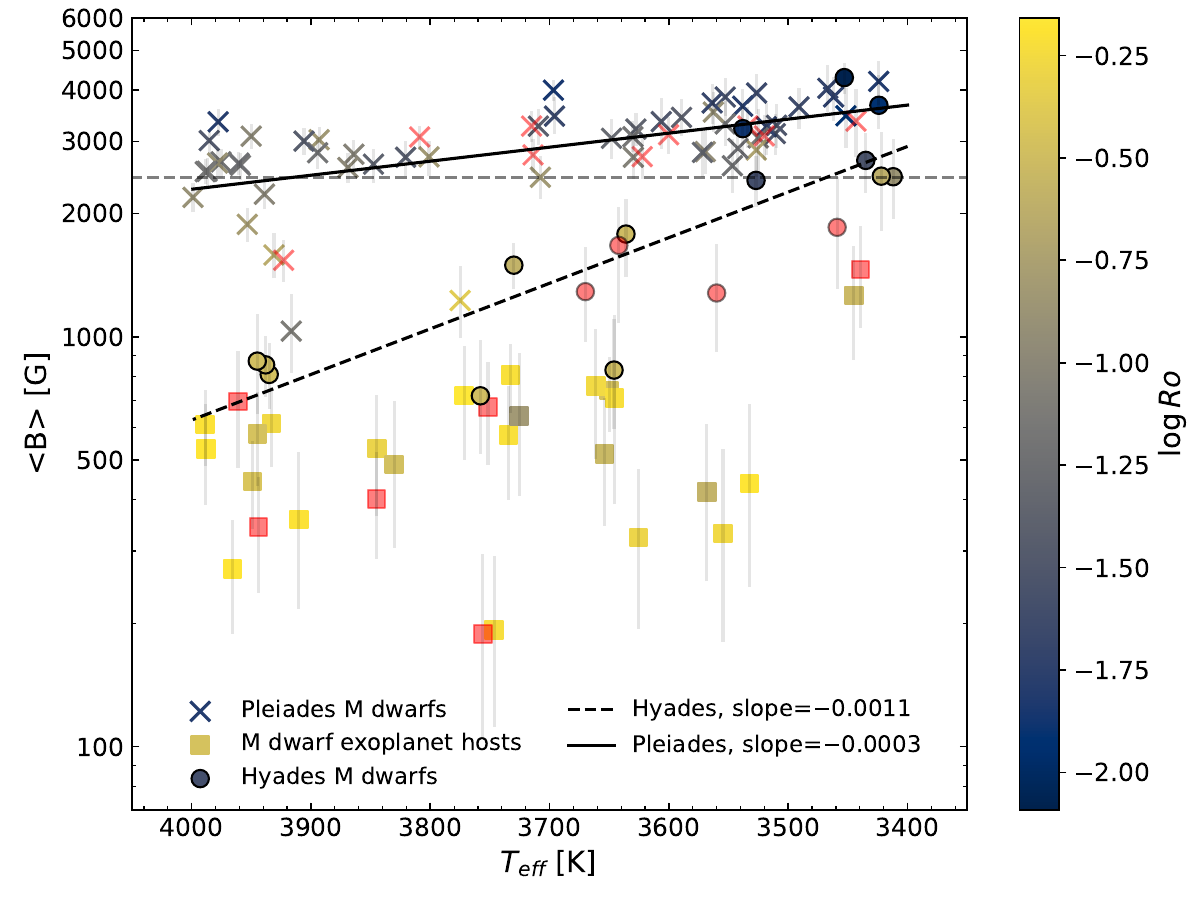}
\caption{Mean magnetic field strengths as a function of effective temperature, color-coded by the logarithm of the Rossby number. Results for the Hyades M dwarfs from this work are shown as circles, while results for Pleiades M dwarfs \citep{wanderley2024} and M dwarfs known to host exoplanets \citep{wanderley2024b} are shown as xs and squares, respectively. Stars in our sample without measured Rossby numbers are highlighted in red. The figure also shows linear regression fits of $\log{\langle \rm B \rangle}$ as a function of $T_{\rm eff}$ for the Pleiades (solid black line) and Hyades (dashed black line) samples. The horizontal dashed gray line indicates the location of $\langle \mathrm{B} \rangle_{\rm SAT}$.}
\end{center}
\label{teffmag}
\end{figure}

Figure \ref{teffmag} shows the relation between mean magnetic fields and effective temperatures for the Pleiades, Hyades, and field star (exoplanet-host) samples, color-coded by the Rossby number (stars without measured rotational periods, or Rossby numbers, from the three samples are colored in red). Linear regression fits of $\log{\langle \rm B \rangle}$ as a function of $T_{\rm eff}$ for the Pleiades and Hyades M dwarfs are represented by the solid and dashed black lines, respectively. 
For the Pleiades M dwarfs, $\log{\langle \rm B \rangle}$ shows a modest dependence on effective temperature, characterized by a best-fit slope of $-0.0003\pm0.0001$. The mean $\pm$ SD of $\langle \rm B\rangle$ increases from $2.41\pm0.61$ kG for Pleiades M dwarfs with $T_{\rm eff}$ between 3900 and 4000 K to $3.76\pm0.30$ kG for those with $T_{\rm eff}$ between 3400 and 3500 K. Most of the Pleiades M dwarfs rotate fast, having Rossby numbers lower than Ro$_{\rm SAT}$. This relatively weak temperature dependence contrasts with the behavior observed in the Hyades M dwarf sample, where hotter stars tend to exhibit significantly weaker magnetic fields. The best-fit slope between $\log\langle \rm B\rangle$ and $T_{\rm eff}$ for the Hyades M dwarfs is $-0.0011\pm0.0002$, substantially steeper than that found for the Pleiades. Furthermore, the difference in mean $\langle \rm B\rangle$ between the 3900--4000 K and 3400--3500 K temperature bins is approximately 2.06 kG for the Hyades sample, compared with 1.35 kG for the Pleiades. This difference is consistent with the Pleiades sample predominantly populating the saturated regime, while the Hyades sample spans both the saturated and unsaturated domains.

In summary, for a given $T_{\rm eff}$, Pleiades M dwarfs possess much stronger magnetic fields than the Hyades stars, reflecting the magnetic field loss in the older Hyades cluster during the age difference that separates the two clusters of $\sim$500 Myr. The five stars with the largest mean magnetic field strengths in the Hyades sample have remarkably lower Rossby numbers than the rest of the sample. These stars are among the coolest in the sample, with $T_{\rm eff}$ values below 3550 K, and their mean magnetic field strengths are very similar to those of Pleiades stars with comparable effective temperatures. This suggests that, at these effective temperatures, the structure of the stars' convective interiors may allow them to retain the bulk of their magnetic fields with little significant loss over the age difference between the two clusters.

For the field stars, we note that the two coolest stars in the sample ($T_{\rm eff}<3450$ K) are also those with $\langle \rm B\rangle>1$ kG. However, overall, there is not a strong relation between $T_{\rm eff}$ and $\log\langle \rm B\rangle$ within the sample. The planetary-host field stars span ages of several Gyr, and their magnetic fields are significantly weaker than those of both clusters at a given $T_{\rm eff}$. This provides further evidence for the gradual weakening of magnetic fields over long timescales in stars with similar convective interiors. We also investigated the relationship between the kinematic ages derived for the field stars and their mean magnetic field strengths, but found no clear correlation, bearing in mind that the age estimates for M dwarfs are subject to large uncertainties. 

\subsection{Mean Magnetic Fields and Stellar Radius Inflation}

Observational studies of M dwarfs have found that they exhibit stellar radii larger than those predicted by stellar isochrone models \citep{torres2002,reiners2012,Jackson2016,jackson2018,Jackson2019,jeffers2018,Kesseli2018,jaehnig2019,cao2025}. This phenomenon, known as radius inflation, is typically quantified as the fractional difference between the observed stellar radius and the radius predicted by the models (R$_{\rm iso}$). Magnetic fields might reduce the efficiency of stellar convection and/or generate stellar spots \citep{chabrier2007}; both effects are responsible for reducing the star's temperature, which causes it to inflate. The previous study by \citet{wanderley2023} investigated radius inflation in the Hyades cluster and found that its M dwarfs are, on average, inflated, although the degree of inflation and its dependence on stellar mass vary depending on the isochrone model employed. 
For the Pleiades, \citet{wanderley2024} found that, on average, Pleiades M dwarfs are more inflated than their Hyades counterparts. These results are overall consistent with our finding that Hyades M dwarfs have, on average, weaker mean magnetic fields than the Pleiades.

\begin{figure*}
\begin{center}  \includegraphics[angle=0,width=1\linewidth,clip]{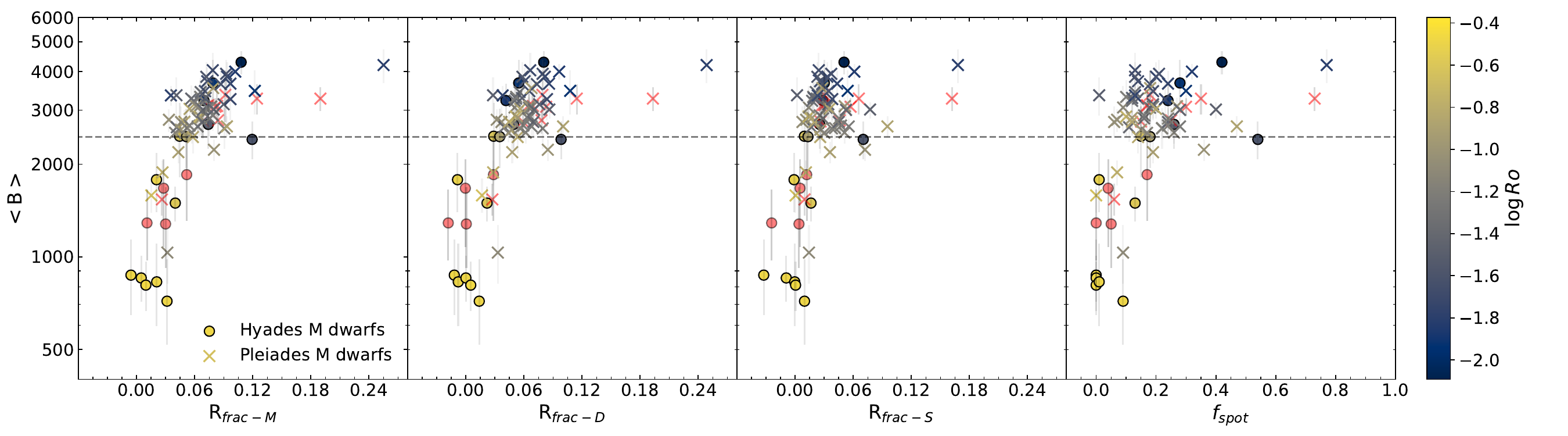}
\caption{The three leftmost panels show mean magnetic field strength as a function of radius inflation, considering, from left to right, the MIST, Dartmouth, and SPOTS isochrones. The rightmost panel shows mean magnetic field strength as a function of spot fractional coverage. Hyades M dwarfs from this work and Pleiades stars from \citet{wanderley2024} are represented by circles and xs, respectively. In both datasets, the points are colour-coded according to their $\log{\mathrm{Ro}}$ values, while objects without available rotational period measurements are shown in red. Gray dashed lines indicate the saturation level, $\langle \mathrm{B} \rangle_{\rm SAT}$.}
\end{center}
\label{rfrac}
\end{figure*}

Given that our stellar parameters in this Hyades study are slightly different from \citet{wanderley2023}, we re-derived radius inflation (R$_{\rm frac}$) for our sample, employing the same methodology as \citet{wanderley2023}. As previously, we selected MIST \citep{choi2016_MIST}, Dartmouth \citep{Dotter2008_dartmouth}, and SPOTS \citep{Somers2020_spots} isochrones with ages of $\sim$0.6 Gyr, and metallicities of $\sim$0.1 as references. Our radius inflation results are in very good agreement with those of \citet{wanderley2023}, with a mean difference of $\sim1\%$ for all three considered isochrone models.

The three leftmost panels of Figure \ref{rfrac} show mean magnetic fields as a function of radius inflation, from left to right, for MIST, Dartmouth, and SPOTS isochrones. The rightmost panel of this figure shows magnetic field strengths as a function of derived fractional spot coverages. Hyades and Pleiades M dwarfs are represented, respectively, as circles and xs. The results are color-coded by their $\log{\mathrm{Ro}}$, while stars without available rotational period measurements are shown in red. The gray dashed lines indicate the position of $\langle \mathrm{B} \rangle_{\rm SAT}$. The figure shows that the Hyades sample follows a pattern that is similar to that of the Pleiades, with stars with greater mean magnetic fields showing higher radius inflation levels and spot coverage fractions, and lower Rossby numbers. 
We note that for one outlier Pleiades star in our sample (2M03443198+2352309), we derived a very large radius inflation and an extremely high spot coverage fraction of 94$\%$, despite this star having a relatively weak magnetic field and a high Rossby number, which places it in the unsaturated regime. \citet{cao2025} identified this star as a candidate binary based on Gaia imaging. Given the potential for flux contamination from an unresolved companion to affect our determinations, we excluded this star from this analysis. 

%While our methodology is based on isochrone fitting, the work of \citet{cao2025} derived fractional spot coverages for stars in open clusters using a two-temperature spectroscopic fitting technique \citep{cao2022}, obtaining a much smaller spot coverage fraction of 15$\%$ for this star. They nevertheless excluded it from their analysis, identifying 
 %the photometric magnitudes used in our luminosity determination, we therefore excluded this star from the present analysis.}

The relationship between R$_{\rm frac}$ and f$_{\rm spot}$ versus $\log{\langle \rm B \rangle}$ mimics the trend shown in Figure \ref{mag}. Stars with weaker magnetic field strengths exhibit a stronger dependence between the two variables, whereas stars with stronger magnetic fields occupy a saturated regime. Stars in the unsaturated regime ($\mathrm{Ro}>\mathrm{Ro}_{\rm SAT}$) exhibit spot coverage fractions of up to 15$\%$, with a mean $\pm$ SD of $6.3\pm5.9\%$. In the saturated regime, the mean $\pm$ SD $f{\rm spot}$ is $21.7\pm12.6\%$. The figure shows that, while there is a clear dependence between mean magnetic fields and radius inflation, the level of inflation depends on the choice of model. 
The mean$\pm$SD inflation levels for the Hyades sample, considering MIST, Dartmouth, and SPOTS isochrones, are respectively 4$\pm$3$\%$ (reaching a maximum of 12$\%$), 2$\pm$3$\%$ (reaching a maximum of 10$\%$), and 1$\pm$2$\%$ (reaching a maximum of 7$\%$). Unlike the MIST and Dartmouth models, for which we only have non-magnetic isochrones for comparison, the SPOTS models allow us to compare isochrones with different fractional spot coverage levels. They consider that stellar spots change the internal structure of the star, suppressing convection \citep{SomersPinsonneault2015}. This affects not only the stellar radii and effective temperature, but also its luminosity, which decreases. Consequently, isochrones with different f$_{\rm spot}$ values exhibit different mass-luminosity relations. This decrease in luminosity results in smaller radius inflation than in models that assume the stellar luminosity remains constant during the inflation process.

%\begin{figure}
%\begin{center}  \includegraphics[angle=0,width=1\linewidth,clip]{fspots.pdf}
%\caption{Mean magnetic fields as a function of fractional spot coverage. Results for Hyades M dwarfs from this work are represented as black circles, and results for Pleiades M dwarfs from \citet{wanderley2024} are represented as gray circles. The gray dashed vertical line represents roughly the separation between the saturated and unsaturated magnetic field regimes for the Hyades sample at f$_{\rm spot}$=15$\%$. The gray horizontal line indicates the position of $\langle \mathrm{B} \rangle_{\rm SAT}$.
%}
%\end{center}
%\label{spots}
%\end{figure}

%\begin{figure*}
%\begin{center}  \includegraphics[angle=0,width=1\linewidth,clip]{emissionlines.pdf}
%\caption{Mean magnetic fields as a function of equivalent width for H$\alpha$ (left panel), H$\beta$ (left panel), and Ca II K (right panel) spectral lines. Results for Hyades M dwarfs from this work are represented as black circles, and results for Pleiades M dwarfs from W24 are represented by gray circles.}
%\end{center}
%\label{emilines}
%\end{figure*}

\begin{figure*}
\begin{center}  \includegraphics[angle=0,width=1\linewidth,clip]{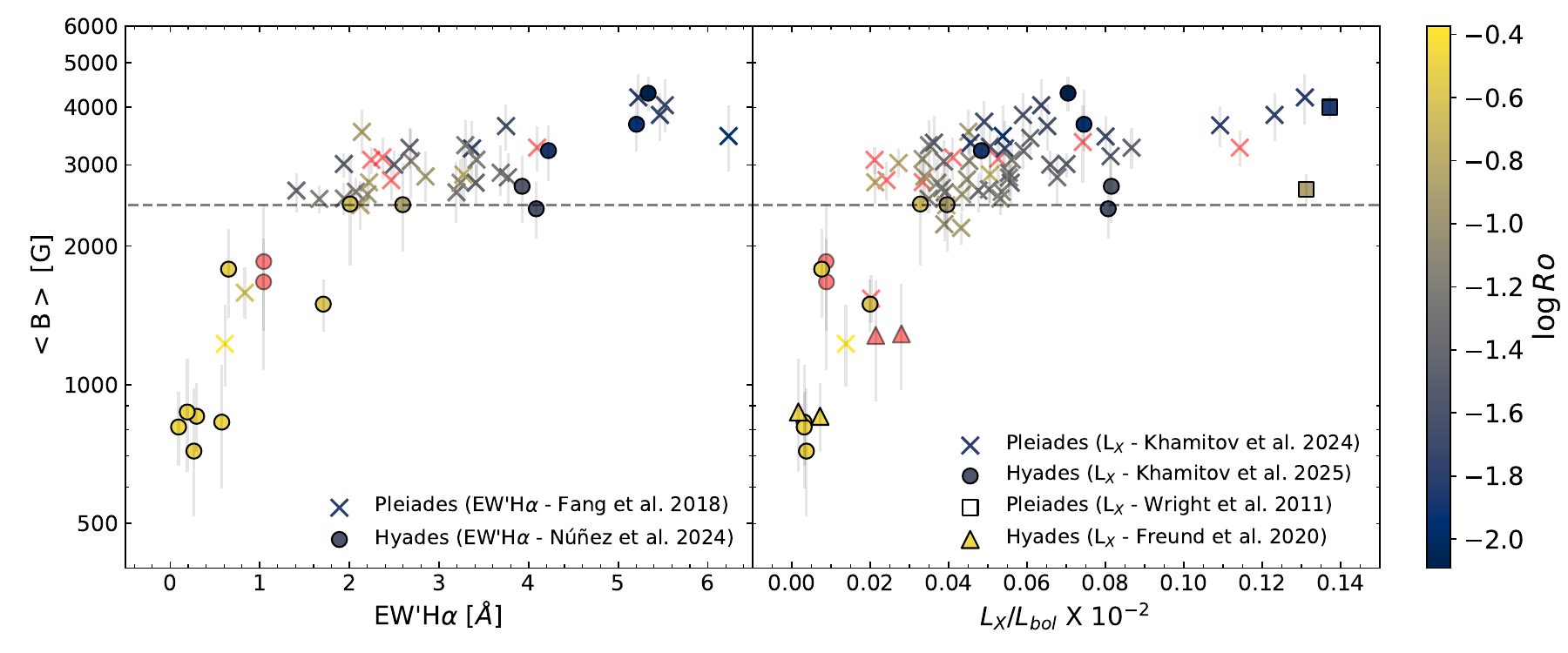}
%\caption{Mean magnetic field strengths as a function of the H$\alpha$ equivalent widths (left panel) and X-ray-to-bolometric luminosity ratios (right panel), color-coded by $\log{\mathrm{Ro}}$. Results for Hyades M dwarfs from this work are shown as circles, while results for Pleiades M dwarfs from \citet{wanderley2024} are shown as xs. Results without rotational period measurements are shown in red. Gray dashed lines indicate the position of $\langle \mathrm{B} \rangle_{\rm SAT}$.}
\caption{Left panel: Mean magnetic field strengths as a function of H$\alpha$ equivalent-width excess (EW'H$\alpha$), color-coded by $\log{\mathrm{Ro}}$. Our results for Pleiades and Hyades stars as a function of H$\alpha$ equivalent width excess from \citet{fang2018} and \citet{nunez2024} are shown as xs and circles, respectively. Right panel: Mean magnetic field strengths as a function of L$_{\rm X}/\rm L_{\rm bol}$, color-coded by $\log{\mathrm{Ro}}$. Circles and xs show our Hyades and Pleiades determinations using X-ray luminosities from \citet{khamitov2025} and \citet{khamitov2024}, respectively. Squares and triangles show our Pleiades and Hyades determinations using X-ray luminosities from \citet{wright2011} and \citet{freund2020}, respectively, for stars without measurements in \citet{khamitov2024,khamitov2025}. For both panels, results without rotational periods are shown in red, while gray dashed lines indicate the position of $\langle \mathrm{B} \rangle_{\rm SAT}$.} 
\end{center}
\label{activity}
\end{figure*}

Both our measurements of radius inflation and stellar spots are highly dependent on the choice of reference model, so it is difficult to determine the real radius inflation and spot coverage of a given star. However, in all cases, the comparison between radius inflation and our derived magnetic fields has shown that, on average, cooler, more magnetic Hyades stars show higher levels of inflation and spot coverage. They also show that the relation between these variables mimics Figure \ref{mag}, which separates the sample into rapid-rotators located in the saturated regime and slow-rotators in the unsaturated regime. These results indicate that starspots associated with strong magnetic fields may contribute to radius inflation in M dwarfs. The derived radius inflation indicators and spot coverage fractions are presented in Table \ref{fulldata}.

\subsection{Mean Magnetic Fields and Activity}

Strong magnetic fields can drive chromospheric activity, leading to enhanced high-energy emission. This activity can manifest as excess ultraviolet emission and as chromospheric spectral lines in emission, such as H$\alpha$ and Ca II H$\&$K. B-fields can also be responsible for heating the stellar corona, inducing X-ray emission, which can be observed as an increase in the X-ray-to-bolometric luminosity ratio. 
%We cross-matched our Hyades M dwarf sample with H$\alpha$ equivalent width excess data from \citet{nunez2024}, finding 16 stars in common. We also compiled X-ray luminosities for Hyades stars from \citet{khamitov2025}, and obtained L$_{\rm X}$/L$_{\rm bol}$ determinations for all the sample by dividing them by our derived bolometric luminosities. For the Pleiades cluster, we compiled H$\alpha$ equivalent width excess and L$_{\rm X}$/L$_{\rm bol}$ data from respectively \citet{fang2018}. and \citet{khamitov2024}.

To investigate the interplay between small-scale magnetic fields and stellar activity indices for the Hyades and Pleiades M dwarfs, in Figure \ref{activity} we show the derived $\langle \mathrm{B} \rangle$ as a function of H$\alpha$ equivalent width excess (in \AA; left panel) and L$_{\rm X}/\rm L_{\rm bol}$ (right panel), color coded according to the logarithm of the Rossby number. For the Hyades sample, we compiled H$\alpha$ equivalent width excess and L$_{\rm X}$/L$_{\rm bol}$ data from \citet{nunez2024} and \citet{khamitov2025}, and for the Pleiades from \citet{fang2018} and \citet{khamitov2024}. 
%Hyades and Pleiades stars are represented by circles and xs, respectively, while stars without available rotational period measurements are shown in red. The magnetic field saturation limit is indicated by gray dashed lines. 
The X-ray luminosities of \citet{khamitov2024,khamitov2025} were derived from eROSITA observations in the 0.3--2.3 keV band, providing the advantage of consistently measured X-ray luminosities for a large number of both Pleiades and Hyades M dwarfs in the same energy range. For those few stars without measurements in these works (two from the Pleiades, four from the Hyades), we used X-ray luminosities from other studies which covered different energy bands: L$_{\rm X}$/L$_{\rm bol}$ from \citet{wright2011} (0.1--2.4 keV) for the Pleiades, and from \citet{freund2020} (0.2--12 keV) for the Hyades.

The distributions of our derived mean magnetic fields as a function of H$\alpha$ equivalent width excess and L$_{\rm X}/\rm L_{\rm bol}$ exhibit a pattern similar to that found in the other relations presented above, with the presence of saturated and unsaturated regimes. Both clusters show evidence of these two regimes, although the Pleiades sample is dominated by stars in the saturated regime, whereas the Hyades sample spans both regimes. In the unsaturated regime, stars with larger Rossby numbers and weaker magnetic fields tend to exhibit lower levels of H$\alpha$ emission and lower X-ray-to-bolometric luminosity ratios. The two panels therefore reveal correlations between the derived mean magnetic field strengths and independent indicators of chromospheric and coronal activity. Together with the Rossby-number dependence, these trends provide evidence for a connection between mean magnetic field strength, stellar rotation, and magnetic activity.

%used LAMOST data to study the chromospheric emission in 4 open clusters, including the Pleiades and Hyades. %They computed excess in equivalent widths of H$\alpha$, H$\beta$, and Ca II K emission lines, by comparing the magnitudes of their studied stars with the ones of inactive field dwarfs. 
%\citet{nunez2024} computed equivalent widths of H$\alpha$ spectral lines for stars from the Praesepe and Hyades open clusters. 
%For the saturated regime, the Hyades stars tend to show lower magnetic fields for the same level of equivalent width when compared to Pleiades stars around roughly equivalent widths from 2 to 4 angstroms. This might suggest that the relationship between chromospheric activity and magnetic fields has a dependence on stellar age. However, our Hyades sample is small within these thresholds, and also the observed offset could be a result of systematic differences between the two equivalent width measurements. To properly assess whether the relationship between magnetic fields and chromospheric activity differs between the two clusters, and to avoid potential biases, a larger and more diverse sample of stars is required.

\section{Conclusions}

Stars within an open cluster share the same age, making different open clusters excellent laboratories for studying how stellar magnetic fields evolve over time. 
The Hyades cluster, with an age of $\sim$600--700 Myr, is particularly valuable in this context. Despite its relatively young age, its M dwarfs span a range of magnetic activity levels, providing an opportunity to investigate the evolution of magnetic fields and their connection to stellar rotation and activity as stars undergo angular momentum loss. 

In this study, we derived small-scale magnetic fields for 18 partially convective M dwarf stars ($3400 < T_{\rm eff} < 4000$ K) from the Hyades open cluster, using near-IR high-resolution spectra from the APOGEE survey. 
%The adopted stellar parameters (T$_{\rm eff}$, $\log{g}$, [M/H], and v$\sin{i}$) were derived employing the same methodology as \citet{wanderley2025}, based on MARCS stellar atmosphere models and the Turbospectrum radiative transfer code. 
These fields were obtained using the same synthetic spectral grid as \citet{wanderley2024}, which was generated with the Synmast radiative transfer code in conjunction with MARCS stellar atmosphere models. We employed an MCMC methodology to infer magnetic filling factors corresponding to field strengths of 2–12 kG by modeling four Zeeman-sensitive Fe I lines. 
Using the same homogeneous methodology, we have derived magnetic fields for three populations of M dwarfs: the Hyades open cluster (this work), the Pleiades open cluster \citep{wanderley2024}, and Galactic field stars hosting exoplanets \citep{wanderley2024b}. The three samples span a similar stellar mass range, allowing us to investigate the dependence of B-field strength on stellar age, rotation, and effective temperature. The distributions of mean magnetic fields for these three M dwarf populations (Pleiades, Hyades and field) reveal clear differences that can be attributed to their different ages (see histograms of Figure \ref{hists}). The Pleiades M dwarfs (ages of $\sim$120 Myr) are approximately 500 million years younger than the Hyades stars, whereas the field M dwarf stars are probably considerably older than both open clusters, having ages in the Gyr range. 
The mean magnetic fields for our Hyades sample range from 0.7 to 4.3 kG, with a mean$\pm$SD of 1.9$\pm$1.0 kG.
The Pleiades M dwarfs exhibit, on average, stronger magnetic fields than their Hyades counterparts, having a mean B-field of $2965\pm632$ G, whereas the field stars exhibit considerably weaker magnetic fields than both cluster samples, with a mean $\pm$ SD of $573\pm272$ G, overall consistent with the expected decline in magnetic activity with increasing stellar age.

We combined the results from the three M-dwarf samples (Hyades, Pleiades, and field M dwarfs) to derive linear relations between $\log{\mathrm{Ro}}$ and $\log{\langle \mathrm{B} \rangle}$ for the saturated and unsaturated regimes, as well as the corresponding saturation limits in Rossby number ($\mathrm{Ro}_{\rm SAT}$) and magnetic field strength ($\langle \mathrm{B} \rangle_{\rm SAT}$). To this end, we used an MCMC approach to simultaneously determine the slopes and intercepts of the two regimes and the Rossby number that separates them, $\mathrm{Ro}_{\rm SAT}$. Each sample from the posterior distribution provides a corresponding value of $\langle \mathrm{B} \rangle_{\rm SAT}$. 
We obtained $\mathrm{Ro}_{\rm SAT}=0.145^{+0.003}_{-0.004}$ and $\langle \mathrm{B} \rangle_{\rm SAT}=2452^{+50}_{-41}$ G. The linear relations for each regime are described in Equations \ref{linearfits2} and \ref{linearfits3}. 

The distribution of mean magnetic field strengths as a function of rotational period and Rossby numbers also reveals distinct patterns among the three samples. As noted above, the Hyades M dwarfs show a large scatter in $\langle \mathrm{B} \rangle$, spanning both the saturated (Ro$\leq$Ro$_{\rm SAT}$) and unsaturated (Ro$>$Ro$_{\rm SAT}$) magnetic regimes. In contrast, the Pleiades M dwarf stars are predominantly in the saturated regime, consistent with their younger ages and greater retention of angular momentum and magnetic fields, and the field M dwarfs are exclusively in the unsaturated regime. (see Figure \ref{mag}).

Stars in the Pleiades and Hyades open clusters show a clear correlation between $\log\langle \mathrm{B} \rangle$ and effective temperature, with cooler stars generally exhibiting stronger mean magnetic fields. However, the relation differs substantially between the two clusters. The Pleiades stars, which are predominantly in the saturated regime, show a considerably flatter relation than the Hyades stars and exhibit stronger magnetic fields at a given $T_{\rm eff}$. The offset in $\langle \mathrm{B} \rangle$ between the two samples is largest among the hottest stars but decreases significantly toward $\sim3400$ K. This suggests that, near this temperature threshold, the age-related differences between the M dwarfs in the two clusters become less pronounced in their magnetic field strengths. This behavior is likely related to the longer spin-down timescales and larger convective envelopes of these cooler stars. For the field-star sample, the relationship between $\log \langle \mathrm{B} \rangle$ and $T_{\rm eff}$ is less clear, reflecting the broader range of stellar ages represented in this sample (see Figure \ref{teffmag}).

We investigated the relationship between the derived small-scale magnetic fields and stellar radius inflation in the Hyades and Pleiades samples, using MIST, Dartmouth, and SPOTS isochrones as baselines. Overall, stars with stronger magnetic fields tended to exhibit greater stellar radius inflation and larger fractional spot coverage. This is consistent with a scenario in which intense magnetic fields generate extensive starspot coverage, which in turn contributes to the observed radius inflation (see Figure \ref{rfrac}). The distribution of radius inflation as a function of mean magnetic field revealed a familiar pattern, characterized by distinct saturated and unsaturated regimes for both the Pleiades and Hyades M dwarfs. The distribution of $f_{\rm spot}$ as a function of $\langle \mathrm{B} \rangle$ likewise follows the same pattern. 

In addition, the relationship between our derived mean magnetic fields for the Pleiades and Hyades samples and measurements of H$\alpha$ equivalent width excess and the X-ray-to-bolometric luminosity ratio reveals clear correlations, with stars exhibiting stronger mean magnetic fields also showing higher levels of chromospheric and coronal activity. In both cases, the distributions display distinct saturated and unsaturated regimes. These results show the close connection between magnetic field strength and stellar activity, consistent with the role of magnetic fields in heating the stellar chromosphere and corona.

The results presented here for the Hyades M dwarfs, alongside comparisons with previous results for M dwarf stars from the Pleiades open cluster and for M dwarfs in the Galactic field, clearly demonstrate the dependence of magnetic fields on both stellar age and mass. Furthermore, they highlight how magnetic field strengths directly correlate with their products, such as stellar radius inflation, star spots, and high-energy emission. Looking forward, expanding these measurements to older open clusters and cooler, fully convective stars is essential to fully constraining the magnetic evolution of M dwarfs.

\clearpage
\startlongtable
\begin{deluxetable*}{cccccccccccccccc}
\small\addtolength{\tabcolsep}{-2pt}
\tablenum{1}
\label{fulldata}
\tabletypesize{\scriptsize}
\tablecaption{Stellar Data}
\tablewidth{0pt}
\tablehead{
\colhead{Sample} &
\colhead{APOGEE ID} &
\colhead{$T_{\rm eff}$} &
\colhead{$v\sin{i}$} &
\colhead{$\langle \rm B \rangle$} &
\colhead{P$_{\rm rot}$} &
\colhead{Ro} &
\colhead{R$_{*}$} &
\colhead{R$_{\rm frac-M}$} &
\colhead{R$_{\rm frac-D}$} &
\colhead{R$_{\rm frac-S}$} &
\colhead{f$_{\rm spot}$} &
\colhead{EW'$_{\rm H \alpha}$} &
\colhead{L$_{\rm X}$/L$_{\rm bol}$} &
\colhead{Age} \\
\colhead{...} &
\colhead{...} &
\colhead{K} &
\colhead{km s$^{-1}$} &
\colhead{G} &
\colhead{day} &
\colhead{...} &
\colhead{R$_{\odot}$} &
\colhead{...} &
\colhead{...}  &
\colhead{...}  &
\colhead{...}  &
\colhead{\AA}  &
\colhead{$\times 10^{-2}$}  &
\colhead{Gyr}  &
}
\startdata
Hyades & 2M03434706+2051363 & 3412 & 3.3 & 2458$_{-517}^{+577}$ & 12.31 & 0.13 & 0.38$\pm0.02$ & 0.05 & 0.04 & 0.01 & 0.18 & 2.60 & 0.04 & ...\\
Hyades & 2M04425217+2320224 & 3422 & 5.3 & 2466$_{-656}^{+700}$ & 21.78 & 0.24 & 0.38$\pm0.02$ & 0.04 & 0.03 & 0.01 & 0.15 & 2.01 & 0.03 & ...\\
... & ... & ... & ... & ... & ... & ... & ... & ... & ... & ... & ... & ... & ... & ... \\
\enddata
\tablenotetext{}{Stellar data for the Hyades (this work), together with a compilation of Pleiades stars from \citet{wanderley2024} and field stars from  \citet{wanderley2024b}. The typical uncertainty in $T_{\rm eff}$ and $v\sin{i}$ for the Hyades sample are respectively $\pm$100 K and $\pm 1~\text{km s}^{-1}$. The complete table is available in electronic format.}
\end{deluxetable*}

\acknowledgments

We thank the referee for comments that improved the paper. 
F.W. acknowledges support from CNPq postdoctoral fellowships at the Universidade Federal do Rio de Janeiro, Observatório do Valongo (process 153388/2025-0), and Uppsala University (process 447495/2025-9). O.K. acknowledges support by the Swedish Research Council (grant agreement no. 2023-03667) and the Swedish National Space Agency.   
K.C. and V.S. acknowledge that their work here is supported, in part, by the National Science Foundation through NSF grant No. AST-2009507. 
D.S. acknowledges support from the Foundation for Research and Technological Innovation Support of the State of Sergipe (FAPITEC/SE) and the National Council for Scientific and Technological Development (CNPq), under grant numbers 794017/2013 and 444372/2024-5.

Funding for the Sloan Digital Sky Survey IV has been provided by the Alfred P. Sloan Foundation, the U.S. Department of Energy Office of Science, and the Participating Institutions. SDSS-IV acknowledges support and resources from the Center for High-Performance Computing at the University of Utah. The SDSS web site is www.sdss.org.
SDSS-IV is managed by the Astrophysical Research consortium for the Participating Institutions of the SDSS Collaboration including the Brazilian Participation Group, the Carnegie Institution for Science, Carnegie Mellon University, the Chilean Participation Group, the French Participation Group, Harvard-Smithsonian Center for Astrophysics, Instituto de Astrof\'isica de Canarias, The Johns Hopkins University, 
Kavli Institute for the Physics and Mathematics of the Universe (IPMU) /  University of Tokyo, Lawrence Berkeley National Laboratory, Leibniz Institut f\"ur Astrophysik Potsdam (AIP),  Max-Planck-Institut f\"ur Astronomie (MPIA Heidelberg), Max-Planck-Institut f\"ur Astrophysik (MPA Garching), Max-Planck-Institut f\"ur Extraterrestrische Physik (MPE), National Astronomical Observatory of China, New Mexico State University, New York University, University of Notre Dame, Observat\'orio Nacional / MCTI, The Ohio State University, Pennsylvania State University, Shanghai Astronomical Observatory, United Kingdom Participation Group,
Universidad Nacional Aut\'onoma de M\'exico, University of Arizona, University of Colorado Boulder, University of Oxford, University of Portsmouth, University of Utah, University of Virginia, University of Washington, University of Wisconsin, Vanderbilt University, and Yale University.

\facility {Sloan}

\software{Matplotlib \citep{Hunter2007_matplotlib}, Numpy \citep{harris2020_numpy}, Turbospectrum \citep{plez2012_turbospectrum}, Synmast \citep{kochukhov2010_synmast}}

\bibliographystyle{yahapj}
\bibliography{references.bib}

\end{document}